\documentclass[10pt,twocolumn,letterpaper]{article}

\usepackage[pagenumbers]{cvpr}
\usepackage{times}
\usepackage{epsfig}
\usepackage{graphicx}
\usepackage{amsmath}
\usepackage{amssymb}
\usepackage{booktabs}
\usepackage{multirow}
\usepackage{array}
\usepackage{caption}
\usepackage{subcaption}
\usepackage{microtype}
\usepackage{xcolor}
\usepackage{url}

\definecolor{cvprblue}{rgb}{0.21,0.49,0.74}
\usepackage[pagebackref,breaklinks,colorlinks,citecolor=cvprblue,linkcolor=cvprblue,urlcolor=cvprblue]{hyperref}

\cvprfinalcopy

\title{CustomDancer: Customized Dance Recommendation\\by Text-Dance Retrieval}

\author{Yawen Qin\textsuperscript{1} \quad Ke Qiu \quad Qin Zhang}

\date{}

\begin{document}
\maketitle
\footnotetext[1]{This work was completed at South-Central Minzu University.}

\begin{abstract}
Dance serves as both a cultural cornerstone and a medium for personal expression, yet the rapid growth of online dance content has made personalized discovery increasingly difficult.
Text-based dance retrieval offers a natural interface for users to search with choreographic intent, but it remains underexplored because dance requires simultaneous reasoning over linguistic semantics, musical rhythm, and full-body motion dynamics.
We introduce \textbf{TD-Data}, a large-scale open dataset for text-dance retrieval, containing about 4,000 12-second dance clips, 14.6 hours of motion, 22 genres, and annotations from professional dance experts.
On top of this dataset, we propose \textbf{CustomDancer}, a multimodal retrieval framework that aligns text with dance through a CLIP-based text encoder, music and motion encoders, and a music-motion blending module.
CustomDancer achieves state-of-the-art performance on TD-Data, reaching 10.23\% Recall@1 and improving retrieval quality in both quantitative benchmarks and user preference studies.
\end{abstract}

\section{Introduction}

\begin{figure}[t]
  \centering
  \includegraphics[width=\linewidth]{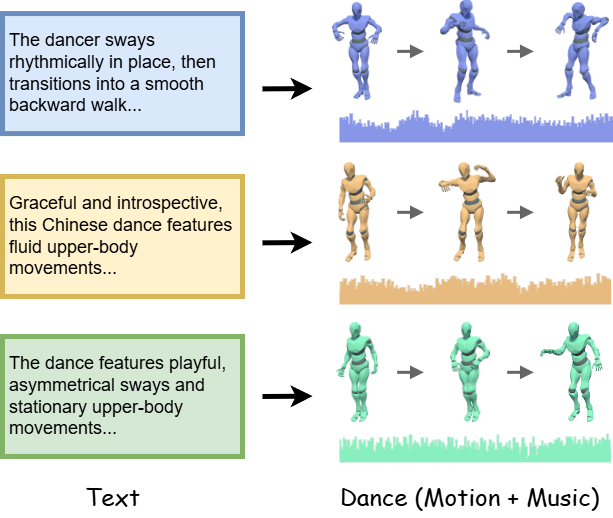}
  \caption{Overview of the text-dance retrieval task. Given a natural-language query, the system retrieves dance clips that match both motion semantics and musical context.}
  \label{fig:task}
\end{figure}

Dance is an ancient and expressive art form with significant cultural and social value.
Each style, from Flamenco to Ballet, encodes history, aesthetics, and community practice through coordinated music and movement.
In modern media platforms, however, the amount of dance content has grown far beyond what users can browse manually.
Search and recommendation interfaces are therefore becoming central to how people discover choreography, learn styles, and create personalized dance experiences.

Most existing retrieval systems are not designed for dance.
Text-to-motion generation methods typically emphasize generic actions such as walking, sitting, or jumping, while overlooking rhythm, style, footwork, and expressive torso movement.
Music-to-dance systems capture beat synchronization but often lack a natural-language interface.
Video-text retrieval models can match high-level events, but they do not explicitly model the body-centric and music-conditioned structure that makes dance distinctive.
This creates a gap between what users want to ask, such as ``a sharp popping sequence with isolated arm movements'', and what current systems can reliably retrieve.

To bridge this gap, we formalize \emph{text-dance retrieval}: given a textual description, retrieve the dance sequence whose music and motion best satisfy the query.
The task is challenging for three reasons.
First, dance descriptions require domain-specific vocabulary.
Second, dance quality depends on multimodal consistency: movement, rhythm, tempo, and style must be interpreted jointly.
Third, public datasets with aligned text, music, and 3D dance motion are scarce.

We address these challenges from both the data and model perspectives.
We construct TD-Data from FineDance~\cite{si2022finedance}, segmenting long motion sequences into coherent clips and enriching each clip with expert annotations and natural-language descriptions.
We then propose CustomDancer, a retrieval model that encodes text, music, and 3D motion in a shared embedding space.
The framework combines CLIP-style language representations~\cite{radford2021clip} with temporal encoders for music and motion, followed by a fusion module that captures complementary and interactive multimodal cues.

Our contributions are summarized as follows:
\begin{itemize}
  \item We benchmark the text-dance retrieval task and introduce TD-Data, a large open dataset with expert-guided text annotations for dance retrieval.
  \item We propose CustomDancer, a multimodal framework that aligns text with music-conditioned 3D dance motion.
  \item We conduct extensive experiments, ablations, user studies, and visual analyses demonstrating the effectiveness of the proposed dataset and method.
\end{itemize}

\section{Related Work}

\subsection{3D Motion Generation}

3D motion generation has evolved from producing short, action-level human movements to modeling long, semantically controlled, and rhythmically structured motion sequences.
Early text-to-motion systems learn correspondences between natural-language descriptions and body kinematics, while recent methods improve realism and controllability through contrastive pretraining, discrete motion tokens, masked modeling, and diffusion priors~\cite{tevet2023motionclip,tevet2022human,petrovich2023tm2t,guo2022generating,zhang2023t2mgpt,guo2024momask,zhang2022motiondiffuse}.
This line of work is important for text-dance retrieval because it shows that language can supervise subtle motion differences rather than only coarse action categories.
HumanML3D-style annotations~\cite{guo2022generating}, T2M-GPT~\cite{zhang2023t2mgpt}, and MoMask~\cite{guo2024momask} are representative examples: they demonstrate that motion features should preserve temporal structure, body-part coordination, and compositional semantics.
Co-speech and skeleton-based motion studies make a similar point from another angle.
SemTalk and EchoMask introduce semantic emphasis and speech-queried mask modeling for holistic co-speech motion~\cite{zhang2025semtalk,zhang2025echomask}; global-rotation diffusion reduces accumulated motion errors under multi-level constraints~\cite{zhang2026globalrotation}; and complexity-aware masked generation allocates modeling capacity according to motion spectral descriptors~\cite{zhou2026dynmask}.
Robust 2D skeleton action recognition via 3D latent distillation further suggests that compact 3D priors can benefit perception even when observations are incomplete~\cite{zhang2025robust2d}.
Together, these works motivate our use of explicit temporal encoders and 3D kinematic features instead of collapsing dance into a single visual embedding.

Dance generation adds another layer of difficulty because motion must remain synchronized with music, style, and choreographic intent.
Music2Dance, Dance Revolution, AIST++, and FineDance establish important foundations for music-conditioned dance synthesis and high-quality 3D dance data~\cite{tang2020music2dance,huang2021dance,li2021aistpp,si2022finedance}.
Subsequent work improves long-range structure and controllability: Bailando and Bailando++ use choreographic memory and GPT-style token prediction for music-to-dance generation~\cite{li2022bailando,siyao2023bailandopp}, while Duolando extends this direction to follower dance accompaniment with off-policy reinforcement learning~\cite{siyao2024duolando}.
Lodge, multi-modal control, InterDance, InfiniteDance, and SoulDance further explore diffusion, control signals, duet interaction, scalable data/model design, and hierarchical motion modeling for music-aligned holistic dance~\cite{li2024lodge,li2024multicontrol,li2024interdance,li2026infinitedance,li2025souldance}.
TM2D combines music and text conditions for 3D dance generation~\cite{gong2023tm2d}.
Recent dance models also move toward retrieval-aware, genre-aware, and efficient architectures.
CoDancers and CoheDancers model coherent group choreography and interactive music-driven decomposition~\cite{yang2024codancers,yang2025cohedancers}; MEGADance introduces a mixture-of-experts design for genre-aware 3D dance generation~\cite{yang2025megadance}; TokenDance formulates token-to-token music-to-dance generation with bidirectional Mamba~\cite{yang2025tokendance}; FlowerDance uses MeanFlow for efficient 3D dance generation~\cite{yang2025flowerdance}; and BiTDiff studies fine-grained 3D conducting motion with BiMamba-Transformer diffusion~\cite{yang2025bitdiff}.
For video-level generation, MACE-Dance first constructs 3D motion and then uses motion-appearance cascaded experts to drive music-conditioned dance video synthesis~\cite{yang2025macedance}, reminding us that motion alignment and appearance synthesis are related but distinct problems.
CustomDancer is not a generator, but these systems clarify the representation requirements for retrieval: a useful dance embedding must retain music-motion synchronization, style, genre, and fine-grained temporal evidence.

\subsection{Multimodal Retrieval}

Multimodal retrieval learns a shared space in which queries from one modality can retrieve content from another.
CLIP popularized large-scale contrastive alignment between language and vision~\cite{radford2021clip}, and similar objectives have since been adapted to audio, video, music, and human motion.
For text-audio matching, AudioCLIP extends image-text contrastive learning to audio~\cite{guzhov2022audioclip}, CLAP aligns natural-language descriptions with acoustic concepts~\cite{huang2023clap}, and natural-language audio retrieval benchmarks show that free-form text queries can support practical sound search~\cite{koepke2022audio}.
In the music domain, MuLan learns a joint embedding between music audio and natural language at large scale~\cite{huang2022mulan}, while CLaMP performs contrastive language-music pretraining for symbolic music retrieval~\cite{wu2023clamp}.
These music retrieval systems are relevant because users often search for affective, rhythmic, and stylistic properties rather than exact labels.
BeatDance further shows that retrieval objectives can be made dance-specific by aligning music and dance through beat-based contrastive learning~\cite{yang2024beatdance}.
For text-video retrieval, XPool and TABLE improve temporal pooling and tag-aware alignment~\cite{gabeur2020multimodal,liu2023table}; for language-conditioned motion retrieval, contrastive learning provides a direct way to compare textual intent with body dynamics~\cite{ghosh2023language}.
These methods are relevant because text-dance retrieval is also a cross-modal ranking problem, but the candidate side is not a generic image or video: it is a synchronized music-motion object.
The model must decide whether a textual phrase such as ``sharp arm isolations over a fast beat'' refers to motion texture, rhythmic placement, musical mood, or their combination.

Existing multimodal retrieval frameworks therefore cannot be transferred to dance without adaptation.
A video-language model may retrieve clips with visually similar scenes while ignoring 3D kinematics, and a music-language model may retrieve rhythmically appropriate audio while missing the described body movement.
Conversely, a pure text-motion model may match action semantics but ignore whether the movement belongs to a musical phrase.
TD-Data and CustomDancer are designed around this gap.
TD-Data supplies aligned descriptions, music, and 3D motion rather than captions alone, and CustomDancer builds the candidate embedding from both acoustic and kinematic streams before contrastive alignment.
In this sense, our work follows the broad multimodal retrieval paradigm while specializing the representation for dance, where semantic relevance is inseparable from temporal coordination and style.

\section{TD-Data Dataset}

\begin{figure*}[t]
  \centering
  \includegraphics[width=0.9\textwidth]{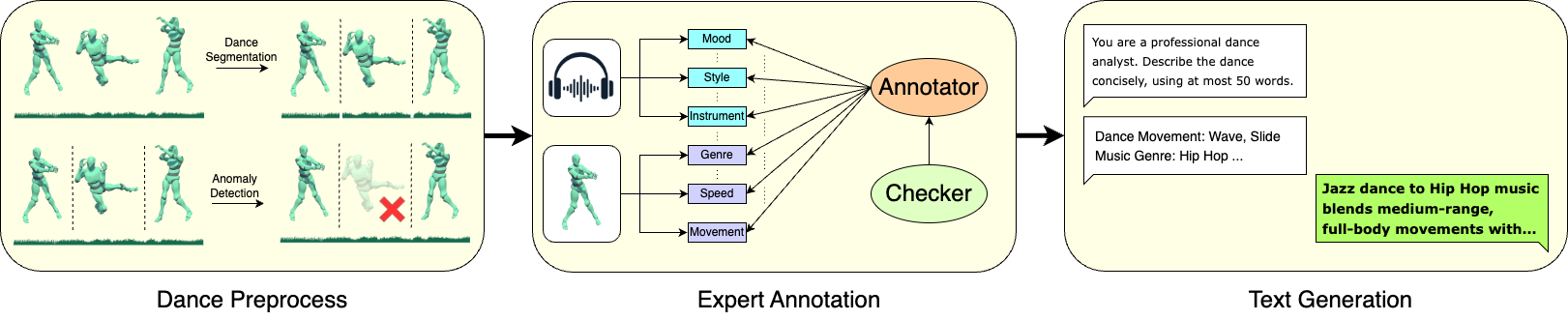}
  \caption{Overview of the TD-Data construction pipeline. Raw dance sequences are segmented, annotated with expert music and motion attributes, validated, and converted into natural-language descriptions.}
  \label{fig:data_pipeline}
\end{figure*}

\subsection{Data Collection}

Text-dance retrieval uses 3D motion data such as SMPL parameters rather than 2D videos because 3D representations explicitly encode spatiotemporal kinematics, including joint rotations, velocities, and body configurations.
These signals allow accurate matching between textual movement descriptions and candidate dance clips.
We derive TD-Data from FineDance and design a three-stage pipeline: dance preprocessing, expert annotation, and AI-assisted text generation.

\subsubsection{Dance Preprocessing}

To balance computational efficiency and semantic completeness, we segment raw motion sequences into 12-second clips.
This duration captures complete dance phrases: a typical eight-beat cycle spans about four seconds, and 12 seconds can contain multiple repetitions, transitions, and expressive variations.
Segmentation also standardizes the retrieval unit, making training batches more consistent and evaluation more interpretable.

\subsubsection{Expert Annotation}

We decompose dance descriptions into a hierarchical taxonomy of music and motion attributes, informed by professional choreographic terminology.
Music attributes include genre, rhythm, tempo, and emotional valence.
Motion attributes include signature movements of the arms, legs, and torso, as well as fluidity, spatial dynamics, and stylistic intensity.

Two certified dance professionals independently annotate each clip.
One acts as the primary annotator and labels the attributes, while the other validates consistency.
Disagreements trigger re-annotation until consensus is reached.
This process improves label reliability and reduces noise in the final text descriptions.

\subsubsection{AI-Assisted Text Generation}

Structured tags are converted into natural-language captions using a controlled prompt to GPT-4o.
The model is instructed to act as a professional dance analyst, describe each dance concisely, vary sentence structure, and preserve choreographic semantics.
The expert-provided attributes remain the source of truth, while the language model helps produce fluent query-like descriptions.

\subsection{Data Statistics}

TD-Data contains about 4,000 high-fidelity 3D dance clips, totaling 14.6 hours at 30 FPS.
The dataset spans 22 genres, including Ballet, Krump, Hip-Hop, Jazz, and other styles, and is performed by 27 professional dancers to reduce individual stylistic bias.
Each clip captures full-body kinematics with 52 joints and is paired with music and a natural-language description.
These properties make TD-Data suitable for evaluating retrieval models that must reason across text, audio, and motion.

\subsection{Annotation Quality and Query Diversity}

The usefulness of a retrieval benchmark depends not only on the number of clips, but also on whether textual queries cover the expressive variation that users actually ask for.
TD-Data therefore includes several complementary description types.
Some captions emphasize style, such as ``a high-energy krump sequence with explosive chest hits''.
Others emphasize rhythm, such as ``a steady hip-hop phrase with repeated accents on the downbeat''.
Additional captions describe body regions, spatial direction, transition quality, and emotional tone.
This diversity discourages shortcut matching based only on genre names and encourages the model to learn fine-grained correspondences between words and motion.

During validation, annotators check whether each caption remains faithful to both motion and music.
Captions that mention absent movements, incorrect tempo, or misleading style terms are rejected and regenerated from the structured tags.
We also preserve compact descriptions rather than verbose summaries, because real search queries tend to be short and selective.
The resulting dataset supports both professional choreographic terminology and practical user-facing retrieval.

\subsection{Split Protocol}

We split TD-Data at the clip level while monitoring performer and genre coverage.
The training set is used to learn cross-modal alignment, the validation set is used for model selection, and the test set is held out for final retrieval evaluation.
Because dance performances from the same dancer can share stylistic signatures, we avoid constructing a test set dominated by a single performer or genre.
This makes the benchmark more faithful to the intended deployment scenario, where a user may search across unfamiliar dancers, songs, and movement vocabularies.

\section{Method}

\begin{figure*}[t]
  \centering
  \includegraphics[width=0.78\textwidth]{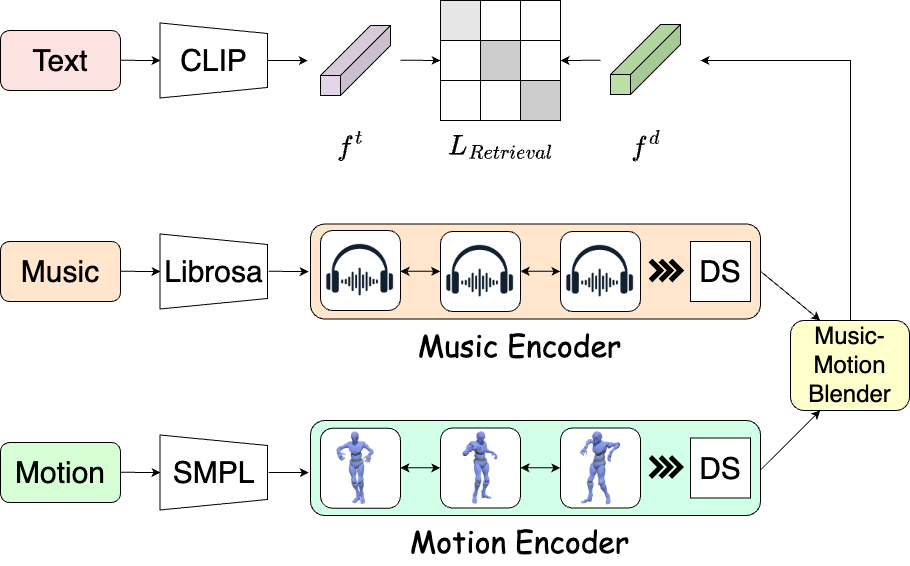}
  \caption{Overview of CustomDancer. Text is encoded by a CLIP-based language module, while music and motion are processed by temporal encoders. The music-motion blender fuses candidate dance features before contrastive alignment with text.}
  \label{fig:framework}
\end{figure*}

\subsection{Problem Definition}

Given a natural-language query $q$ and a gallery of dance candidates $\mathcal{D}=\{d_i\}_{i=1}^{N}$, text-dance retrieval aims to rank candidates so that the semantically matched dance appears near the top.
Each dance candidate consists of synchronized music $a_i$ and 3D motion $m_i$.
The goal is to learn a scoring function
\begin{equation}
  s(q,d_i)=\mathrm{sim}\bigl(f_t(q), f_d(a_i,m_i)\bigr),
\end{equation}
where $f_t$ is the text encoder, $f_d$ is the dance encoder, and $\mathrm{sim}(\cdot,\cdot)$ denotes cosine similarity.

\subsection{Framework Overview}

CustomDancer models the triadic relationship among text, music, and motion using four modules:
a text encoder, a music encoder, a motion encoder, and a music-motion blender.
The text encoder extracts semantic features from the query.
The music encoder captures temporal-acoustic patterns such as rhythm, timbre, and onset structure.
The motion encoder models full-body dynamics from 3D pose sequences.
The blender combines music and motion into a unified dance representation for cross-modal retrieval.

For text input, the latent representation is
\begin{equation}
  \mathbf{z}_t = f_t(q) \in \mathbb{R}^{d}.
\end{equation}
For music and motion, the model produces temporal embeddings
\begin{equation}
  \mathbf{H}_a = f_a(a) \in \mathbb{R}^{T_a \times d}, \qquad
  \mathbf{H}_m = f_m(m) \in \mathbb{R}^{T_m \times d}.
\end{equation}
The final dance embedding is obtained by blending and pooling these streams:
\begin{equation}
  \mathbf{z}_d = \mathrm{Pool}\bigl(g(\mathbf{H}_a,\mathbf{H}_m)\bigr).
\end{equation}

\subsection{Text Encoder}

We initialize the text encoder with the pre-trained text transformer from CLIP~\cite{radford2021clip}.
CLIP's language representation is effective for cross-modal alignment and provides a strong prior for mapping descriptive phrases into a semantic embedding space.
Because dance retrieval requires domain adaptation, we add a lightweight two-layer MLP adapter:
\begin{equation}
  \mathbf{z}_t = \mathrm{MLP}\bigl(\mathrm{CLIPText}(q)\bigr),
\end{equation}
where $\mathrm{MLP}: \mathbb{R}^{d_c}\rightarrow\mathbb{R}^{d}$ projects CLIP embeddings into the retrieval space.

\subsection{Music Encoder}

We extract 35-dimensional Librosa features~\cite{mcfee2015librosa}, including MFCC delta coefficients, chroma features, and onset descriptors.
These features are selected because they capture dance-relevant musical cues such as tempo, rhythmic accents, and harmonic changes.
To model long-range dependencies, the sequence is passed through stacked Transformer encoders~\cite{vaswani2017attention} interleaved with one-dimensional convolutional downsampling:
\begin{equation}
  \mathbf{H}_a^{\ell+1} =
  \mathrm{Down}\bigl(\mathrm{Transformer}_{\ell}(\mathbf{H}_a^{\ell})\bigr).
\end{equation}
Inside each Transformer layer, temporal dependencies are computed with scaled dot-product attention:
\begin{equation}
  \mathrm{Attention}(Q,K,V)
  =
  \mathrm{softmax}\left(\frac{QK^{\top}}{\sqrt{d_k}}\right)V,
  \label{eq:attention}
\end{equation}
where $Q$, $K$, and $V$ are query, key, and value projections of temporal music tokens, and $d_k$ is the key dimension used for scale normalization.
This operation lets every music frame attend to distant rhythmic events, which is important when a dance phrase responds to earlier beats or repeated musical motifs.
Each downsampling layer uses a kernel size of 3 and stride 2, reducing temporal resolution while preserving local continuity.

\subsection{Motion Encoder}

We represent 3D dance motion using SMPL parameters~\cite{loper2015smpl}, which explicitly model body shape and pose.
Given a motion sequence $\mathbf{M}\in\mathbb{R}^{T\times p}$, the encoder maps it into a temporal feature sequence:
\begin{equation}
  \mathbf{H}_m = f_m(\mathbf{M}).
\end{equation}
The motion encoder uses alternating Transformer blocks and downsampling layers.
Self-attention captures global interaction across distant frames, for example correlating preparatory poses with later jumps or spins.
Downsampling compresses the sequence from $T$ to approximately $T/8$, emphasizing semantically salient patterns while reducing computational cost.

\subsection{Music-Motion Blender}

The music-motion blender fuses candidate dance features using both additive and multiplicative interactions:
\begin{equation}
  \mathbf{B} =
  \phi\left(
  W\left[
  \mathbf{H}_a \oplus \mathbf{H}_m;\;
  \mathbf{H}_a \otimes \mathbf{H}_m
  \right]\right),
\end{equation}
where $\oplus$ denotes element-wise addition, $\otimes$ denotes the Hadamard product, $W$ is a learnable projection, and $\phi$ is a nonlinear activation.
Temporal average pooling aggregates the fused sequence:
\begin{equation}
  \mathbf{z}_d = \frac{1}{T}\sum_{t=1}^{T}\mathbf{B}_t.
\end{equation}
The additive path preserves complementary information, while the multiplicative path highlights cross-modal agreement.

\subsection{Training Objective}

We adapt the CLIP contrastive loss for unidirectional text-to-dance alignment.
For a text feature $f_i^{\mathrm{text}}$ and its matched dance feature $f_i^{\mathrm{dance}}$, the per-sample objective is
\begin{equation}
  \mathcal{L}_i
  =
  -\log
  \left[
  \frac{
  \exp(\mathrm{sim}(f_i^{\mathrm{text}},f_i^{\mathrm{dance}})/\tau)}
  {\sum_{j}\exp(\mathrm{sim}(f_i^{\mathrm{text}},f_j^{\mathrm{dance}})/\tau)}
  \right],
  \label{eq:sample_loss}
\end{equation}
where the numerator scores the positive text-dance pair and the denominator contrasts it against all dance candidates in the mini-batch.
For a batch of $B$ matched text-dance pairs, the training loss averages Eq.~\ref{eq:sample_loss}:
\begin{equation}
  \mathcal{L}
  =
  -\frac{1}{B}\sum_{i=1}^{B}
  \log
  \frac{\exp(\mathrm{sim}(\mathbf{z}_{t,i},\mathbf{z}_{d,i})/\tau)}
  {\sum_{j=1}^{B}\exp(\mathrm{sim}(\mathbf{z}_{t,i},\mathbf{z}_{d,j})/\tau)},
  \label{eq:loss}
\end{equation}
where $\tau$ is a learnable temperature parameter.
This objective strengthens the joint embedding space while accommodating the asymmetry between user text queries and candidate dance clips.

\subsection{Implementation Details}

All modalities are projected into the same embedding dimension before contrastive training.
Text embeddings are initialized from CLIP and fine-tuned with a smaller learning rate than the newly initialized adapters.
Music and motion encoders are trained from scratch because their input statistics differ substantially from image-language pretraining.
For temporal encoders, we use positional embeddings before the first Transformer block and preserve temporal order through each downsampling stage.

We normalize both text and dance embeddings before computing cosine similarity.
The learnable temperature is initialized to a moderate value to avoid early over-confidence.
In practice, the model benefits from balanced batches that contain varied genres and performers, since homogeneous batches reduce the number of meaningful negatives.
We also apply lightweight dropout inside the MLP adapters and Transformer blocks to improve robustness.

\subsection{Why Multimodal Fusion Matters}

Dance cannot be reliably retrieved from motion alone or music alone.
Motion alone may identify a jump, turn, or arm wave, but miss whether the phrase is relaxed, explosive, syncopated, or lyrical.
Music alone may identify tempo and mood, but cannot distinguish between choreographic patterns performed to similar beats.
The music-motion blender is therefore designed to preserve both complementary and interactive evidence.
Additive fusion keeps information that appears in only one modality, while multiplicative fusion highlights synchronized cues, such as strong body accents aligned with percussion.
This is especially important for dance phrases where the same motion vocabulary can communicate different intent under different musical contexts.

\section{Experiments}

\subsection{Evaluation Metrics}

We evaluate retrieval using Recall@K, Median Rank, and Mean Rank.
Recall@K measures the proportion of queries for which the correct dance appears in the top $K$ retrieved results.
Median Rank and Mean Rank summarize the ranking position of the ground-truth candidate, where lower values indicate better retrieval.
These metrics are widely used in cross-modal retrieval and provide complementary views of precision and ranking quality.

\subsection{Retrieval Protocol}

For each text query in the test set, the model ranks all candidate dance clips according to cosine similarity in the learned embedding space.
The positive candidate is the clip paired with the query through the TD-Data annotation pipeline, while all other clips in the gallery serve as negatives during evaluation.
This setting is deliberately stricter than genre classification: many negatives may share the same genre, tempo, or performer, so the model must rely on finer evidence such as body-part emphasis, movement quality, and rhythm-motion coupling.
We report results over the full test gallery rather than over small candidate subsets, because real retrieval systems must operate under large and visually similar collections.

During training, in-batch negatives are constructed from diverse performers and genres whenever possible.
This reduces the chance that the model learns a shortcut such as matching only to genre words or dancer identity.
We also keep the text-to-dance direction as the primary objective because it matches the intended user interaction: a user types a natural-language query and expects a ranked list of dances.
The reverse direction is useful diagnostically, but it is less central to the recommendation scenario studied in this paper.

\subsection{Comparison with Existing Methods}

\begin{table}[t]
  \centering
  \caption{Performance comparison on the text-dance retrieval task. Higher Recall is better; lower rank is better.}
  \label{tab:comparison}
  \setlength{\tabcolsep}{3.8pt}
  \resizebox{\linewidth}{!}{%
  \begin{tabular}{lccccc}
    \toprule
    Method & R@1$\uparrow$ & R@5$\uparrow$ & R@10$\uparrow$ & MedR$\downarrow$ & MnR$\downarrow$ \\
    \midrule
    TABLE~\cite{liu2023table} & 8.70 & 34.52 & 47.83 & 12.0 & 23.34 \\
    XPool~\cite{gabeur2020multimodal} & 9.46 & 34.27 & 47.57 & 11.0 & 22.50 \\
    CustomDancer & \textbf{10.23} & \textbf{34.78} & \textbf{48.34} & \textbf{11.0} & \textbf{22.09} \\
    \bottomrule
  \end{tabular}
  }
\end{table}

We compare CustomDancer against two strong cross-modal retrieval baselines: XPool~\cite{gabeur2020multimodal}, which aligns text and video through attention-based temporal pooling, and TABLE~\cite{liu2023table}, a tagging-enhanced multimodal retrieval approach.
As shown in Table~\ref{tab:comparison}, CustomDancer achieves the best performance across all Recall metrics and Mean Rank.
The result indicates that explicit music-motion modeling is beneficial for dance retrieval, where the candidate representation must capture both auditory and kinematic semantics.

\subsection{Ablation Study}

\begin{table}[t]
  \centering
  \caption{Effect of temporal modeling architecture.}
  \label{tab:temporal}
  \setlength{\tabcolsep}{4.2pt}
  \resizebox{\linewidth}{!}{%
  \begin{tabular}{lccccc}
    \toprule
    Method & R@1$\uparrow$ & R@5$\uparrow$ & R@10$\uparrow$ & MedR$\downarrow$ & MnR$\downarrow$ \\
    \midrule
    RNN & 6.40 & 23.79 & 41.43 & 14.0 & 27.93 \\
    LSTM & 7.67 & 30.17 & 43.48 & 14.0 & 24.36 \\
    CustomDancer & \textbf{10.23} & \textbf{34.78} & \textbf{48.34} & \textbf{11.0} & \textbf{22.09} \\
    \bottomrule
  \end{tabular}
  }
\end{table}

\begin{table}[t]
  \centering
  \caption{Effect of feature fusion strategy.}
  \label{tab:fusion}
  \setlength{\tabcolsep}{4.2pt}
  \resizebox{\linewidth}{!}{%
  \begin{tabular}{lccccc}
    \toprule
    Method & R@1$\uparrow$ & R@5$\uparrow$ & R@10$\uparrow$ & MedR$\downarrow$ & MnR$\downarrow$ \\
    \midrule
    MUL & 4.86 & 23.27 & 38.62 & 17.0 & 34.72 \\
    ADD & 9.71 & 30.95 & 46.55 & 12.0 & 21.28 \\
    CustomDancer & \textbf{10.23} & \textbf{34.78} & \textbf{48.34} & \textbf{11.0} & 22.09 \\
    \bottomrule
  \end{tabular}
  }
\end{table}

We first ablate the temporal modeling architecture.
Replacing the Transformer backbone with RNN or LSTM encoders substantially degrades performance, as shown in Table~\ref{tab:temporal}.
This confirms the value of global self-attention for dance, where distant frames can be semantically related through preparation, repetition, and release.

We further evaluate feature fusion strategies in Table~\ref{tab:fusion}.
Pure multiplication performs poorly because it overemphasizes shared activations and suppresses complementary cues.
Pure addition is stronger, but it cannot explicitly model cross-modal agreement.
The full blender combines both interactions and obtains the best Recall@1 and Recall@10, demonstrating that dance retrieval benefits from preserving complementarity while still modeling interaction.

\subsection{User Study}

\begin{table}[t]
  \centering
  \caption{Human preference comparison. TMC denotes text-motion consistency, and TMR denotes text-music relevance.}
  \label{tab:user}
  \begin{tabular}{lcc}
    \toprule
    Method & TMC$\uparrow$ & TMR$\uparrow$ \\
    \midrule
    Ground Truth & \textbf{4.43} & \textbf{4.42} \\
    TABLE~\cite{liu2023table} & 3.55 & 3.31 \\
    XPool~\cite{gabeur2020multimodal} & 3.69 & 3.52 \\
    CustomDancer & 3.82 & 3.68 \\
    \bottomrule
  \end{tabular}
\end{table}

To evaluate real-world alignment between retrieved dances and text queries, we conducted a single-blind user study with 10 participants, including amateur dancers, choreographers, and instructors.
Participants rated top-1 retrieval results using five-point Likert scales for text-motion consistency and text-music relevance.
As shown in Table~\ref{tab:user}, CustomDancer outperforms both retrieval baselines and narrows the gap to ground-truth matches.
The improvement suggests that the learned dance embedding better reflects human judgments of choreographic fit.

\subsection{Failure Cases}

Although CustomDancer improves retrieval quality, several failure modes remain.
The first occurs for highly specialized dance terminology.
If a query contains rare professional terms that appear sparsely in TD-Data, the text encoder may map them near more common neighboring styles.
The second occurs when visual motion and musical affect conflict.
For example, a clip may contain sharp movements over soft music, making it ambiguous whether the query should prioritize motion texture or audio mood.
The third failure mode is performer bias: some dancers consistently execute movements with distinctive personal style, and a model can occasionally use that style as a proxy for genre.

These cases suggest two practical improvements.
First, the annotation vocabulary should continue expanding toward expert-level terminology while preserving natural query phrasing.
Second, retrieval interfaces should support interactive refinement, allowing users to add constraints such as tempo, body part, genre, or emotional valence after seeing the first results.
The present model provides a foundation for such systems, but user feedback can further disambiguate intent.

\subsection{Visualization}

\begin{figure*}[t]
  \centering
  \includegraphics[width=0.8\textwidth]{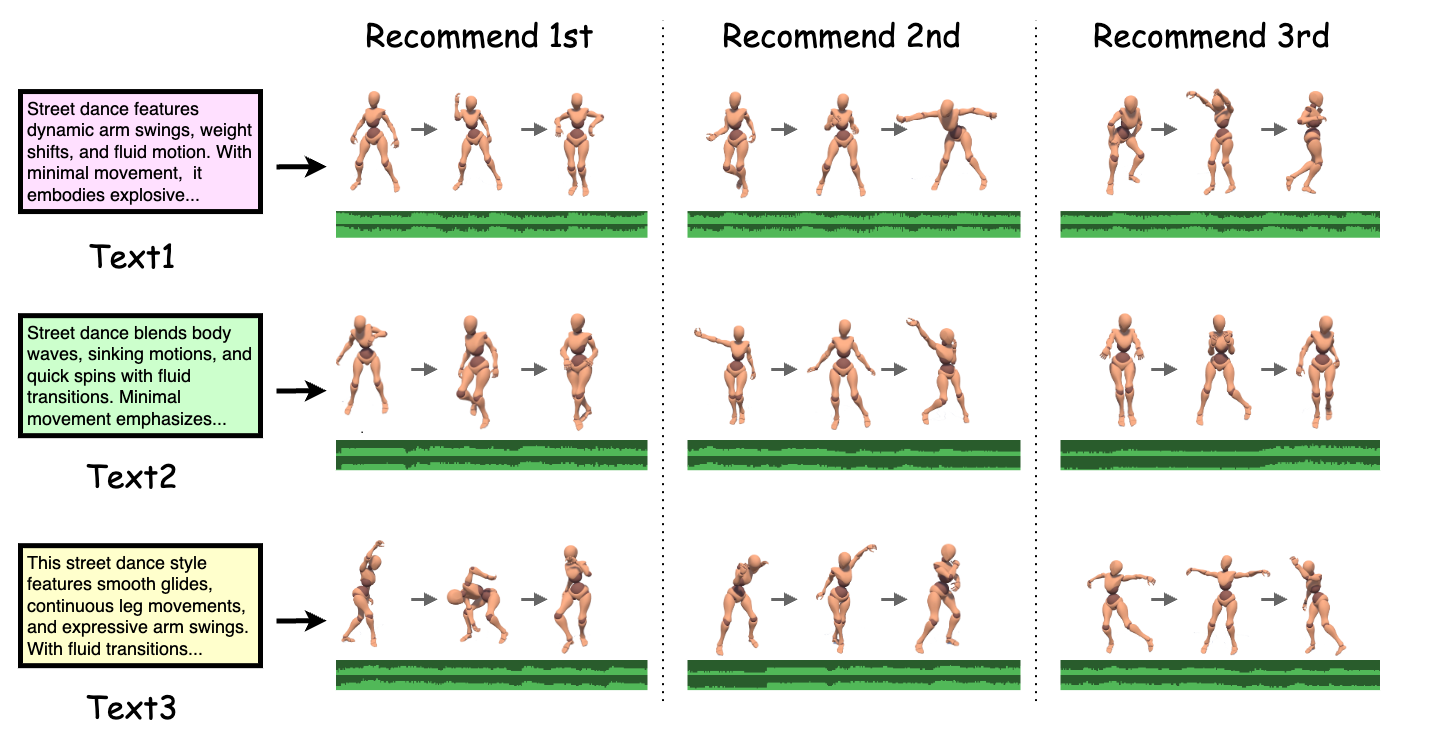}
  \caption{Qualitative retrieval examples from CustomDancer. The examples show that the model can retrieve stylistically and rhythmically aligned dance clips for varied textual queries.}
  \label{fig:visualization}
\end{figure*}

Figure~\ref{fig:visualization} provides qualitative evidence of CustomDancer's text-to-dance retrieval behavior.
Across representative queries, the retrieved results show strong correspondence between described movement style and candidate dance dynamics.
The model is able to distinguish subtle semantic cues, such as sharp versus fluid motion and isolated gestures versus full-body movement.
These examples complement the quantitative results and indicate that the learned embedding captures useful choreographic structure.

The examples also illustrate why text-dance retrieval should be evaluated qualitatively in addition to using rank-based metrics.
Two candidate clips may both contain the correct high-level style, but differ in the body part emphasized, the intensity of movement, or the temporal relationship to the beat.
Human viewers tend to notice these distinctions immediately.
By showing retrieved clips side by side with the query, qualitative visualization helps diagnose whether the model has learned genuine choreographic semantics or merely broad genre correlation.

\subsection{Discussion}

The experimental results suggest that text-dance retrieval is not simply a smaller variant of text-video retrieval.
Dance clips contain repetitive and symmetric patterns, and many visually different movements can satisfy the same high-level description.
At the same time, many visually similar movements differ in choreographic meaning because of timing, intensity, or musical context.
This creates a ranking problem where the correct answer is often separated from plausible negatives by subtle cues.
CustomDancer addresses this by building the candidate embedding from both music and motion, but the benchmark also shows that current retrieval accuracy remains far from saturated.

The ablation results help explain where the remaining difficulty lies.
Temporal modeling has a large effect because dance semantics often unfold across a phrase rather than in a single frame.
Feature fusion also matters because music and motion contribute different types of evidence: music provides tempo, affect, and rhythmic accents, while motion provides body configuration, spatial dynamics, and stylistic texture.
The user study confirms that improvements in retrieval metrics correspond to perceptible differences for human viewers, but it also reveals a gap between model ranking and expert judgment.
Closing this gap will likely require richer text supervision, stronger temporal alignment objectives, and user-controllable retrieval interfaces that can resolve ambiguity after the first ranked results.

\section{Conclusion}

We presented CustomDancer, a multimodal framework for text-dance retrieval, together with TD-Data, a large-scale dataset for aligning natural-language descriptions with music-conditioned 3D dance motion.
By combining a CLIP-based text encoder, temporal music and motion encoders, and a music-motion blending module, CustomDancer effectively models the semantic and rhythmic structure required for dance search.
Experiments show that the proposed method improves retrieval performance over strong cross-modal baselines, and user studies confirm that the retrieved dances better match human judgments.

Future work may extend TD-Data with richer multilingual annotations, more fine-grained choreographic labels, and interactive retrieval feedback.
Another promising direction is to couple retrieval with generation, allowing users to first retrieve relevant dances and then adapt them to new music, style constraints, or performer identities.

\section*{Limitations and Broader Impact}

TD-Data and CustomDancer are designed to make dance search more accessible, but they should be used with attention to cultural context.
Dance styles often carry community-specific history, and reducing them to labels can obscure that context.
Dataset construction should therefore involve domain experts and, where appropriate, practitioners from the represented styles.
The current dataset focuses on 3D motion and music features rather than full video appearance, which avoids some visual privacy concerns but does not fully capture costume, stage setting, facial expression, or camera motion.

The model can support education, choreography browsing, and creative recommendation, but it should not be treated as an authority on cultural authenticity.
Future systems should expose uncertainty, allow users to inspect multiple candidates, and provide transparent metadata about style, performer, and annotation source.
These considerations are especially important if text-dance retrieval is deployed in public creative platforms.

\clearpage

{\small
\bibliographystyle{unsrtnat}
\bibliography{references}
}

\end{document}